%% file: NogatzSeipel.tex
\documentclass[submission,copyright,creativecommons]{eptcs}
\providecommand{\event}{WLP 2016} 
\usepackage{breakurl}             

\title{Implementing GraphQL as a Query Language\\ for Deductive Databases in SWI--Prolog\\ Using DCGs, Quasi Quotations, and Dicts}
\author{Falco Nogatz \quad\quad Dietmar Seipel
\institute{University of W\"urzburg, Department of Computer Science,\\
Am Hubland, D\,--\,97074 W\"urzburg, Germany\\
\email{\{falco.nogatz,dietmar.seipel\}@uni-wuerzburg.de}}
}
\def\titlerunning{Implementing GraphQL as a Query Language for Deductive Databases}
\def\authorrunning{F. Nogatz \& D. Seipel}

\usepackage{xspace}
\usepackage{graphicx}
\usepackage{fancyvrb}
\usepackage{xcolor}

\usepackage{listings}
\definecolor{shadecolor}{RGB}{150,150,150}
\newcommand{\mylabel}[1]{\hfill{\color{darkgray}{\sffamily \colorbox{gray!30}{\textsc{#1}}\strut}\hspace{-0.1cm}}}

\input{SeipelStyles.tex}

\begin{document}
\maketitle

\begin{abstract}
The methods to access large relational databases in a distributed system are well established: the relational query language \sql often serves as a language for data access and manipulation, and in addition public interfaces are exposed using communication protocols like REST. Similarly to REST, GraphQL is the query protocol of an application layer developed by Facebook. It provides a unified interface between the client and the server for data fetching and manipulation. Using GraphQL's type system, it is possible to specify data handling of various sources and to combine, e.g., relational with NoSQL databases. In contrast to REST, GraphQL provides a single API endpoint and supports flexible queries over linked data.

GraphQL can also be used as an interface for deductive databases. In this paper, we give an introduction of GraphQL and a comparison to REST. Using language features recently added to SWI--Prolog 7, we have developed the Prolog library \textit{GraphQL.pl}, which implements the GraphQL type system and query syntax as a domain--specific language with the help of definite clause grammars~(DCG), quasi quotations, and dicts. Using our library, the type system created for a deductive database can be validated, while the query system provides a unified interface for data access and introspection.
\end{abstract}

\section{Introduction}

In times of big data, the increasing growth of data sets is a big challenge for both data storage and data access. The traditional back--end services became more diverse in the recent past. Today, relational databases are used side by side with document--oriented NoSQL databases, each with a different data query mechanism. To bridge over the various query languages, abstraction layers have been introduced.

In October 2015, Facebook open--sourced \textit{GraphQL}~\cite{facebook2015graphql}, an application layer query language, which has been internally developed since 2012. Today, most of Facebook's applications make use of GraphQL as the data--fetching mechanism, resulting in hundreds of billions of GraphQL API calls a day~\cite{GraphQL_Announcement}.

The need for GraphQL arose while developing native mobile applications side by side with web applications. In order to share the code base where possible, a unified application programming interface~(API) for data access should be provided. At the same time, the introduction of an abstraction layer hides the internals like the used database and schemas. In an agile development cycle, this allows to change front--end and back--end technologies simultaneously without impeding each other.

Because of various problems and deficiencies in the alternatives of such an abstraction layer and for accessing hierarchical data, a new query language has been defined. The most popular alternative architectural style REST~\cite{fielding2000architectural}, the representational state transfer, appeared unsuitable for applications that need a lot of flexibility when accessing the data while reducing the number of round--trip times in a distributed system. As a result, GraphQL provides only a single API endpoint for data access, which is backed by a structured, hierarchical type system. Because of this, the GraphQL specification defines besides the query language mechanisms for query validation and provides self--descriptiveness by introspection.

This makes GraphQL a good fit for a unified abstraction layer for static, predefined deductive database queries. The type system is storage--independent and can easily integrate various data sources, like relational and document--oriented databases as well as file--based data formats. With the deductive database system \ddbase~\cite{seipel2015knowledge}, we already have available a system to connect various storage back--ends like \csv and \sql. It is part of the Dislog's Developer Toolkit \ddk \footnote{\url{http://www1.pub.informatik.uni-wuerzburg.de/databases/ddbase/}}, which also contains tools for querying, updating and transforming \xml data. With GraphQL, this results in a interface, which can be integrated in existing back--end environments and interact with any programming language used for the front--end development.

In this paper, we choose to use SWI--Prolog to develop a GraphQL server implementation, called \textit{GraphQL.pl}. With the use of definite clause grammars (DCG), Prolog gives a natural approach to implement the type system and query document as an external domain--specific language so that the code examples used in the GraphQL specification can be used directly to define the GraphQL server. With quasi quotations, which were introduced in SWI--Prolog of version 6.3.17, this domain--specific language can be directly embedded into normal Prolog code, which lowers the barrier for GraphQL--experienced developers who are not yet familiar with Prolog. For the sake of a shorter and prettier syntax, which is more familiar to developers coming from other programming languages, our implementation makes great use of features supported only by recent SWI--Prolog releases. Therefore \textit{GraphQL.pl} requires SWI--Prolog of at least version~7.

The paper is organised as follows: Section~\ref{sec:graphql} introduces GraphQL and gives a comparison to the well--established architectural pattern REST. In Section~\ref{sec:dsl}, we discuss the definition of GraphQL as a domain--specific language in Prolog. Implementation details of \textit{GraphQL.pl} are presented in Section~\ref{sec:implementation}, with a focus on the definition of GraphQL's type system using definite clause grammars, quasi quotations, and dicts. Section~\ref{sec:example} presents a use case of our implementation to define a standalone GraphQL server. Finally, we conclude with a summary and discussion of future work in Section~\ref{sec:conclusion}.

\section{Background: The GraphQL Application Layer}
\label{sec:graphql}

Although used several years in production, the GraphQL specification is still under active development. The most recent Working Draft specification\cite{facebook2015graphql} is of April 2016. Most of the changes since its initial publication in October 2015 have been clarifications in the wording.

Along with the GraphQL specification, Facebook published a reference implementation of a GraphQL server using JavaScript\footnote{\label{note_graphqljs}Facebook on GitHub: GraphQL.js, \url{https://github.com/graphql/graphql-js}}. This reference implementation only provides base libraries that would be the basis for a GraphQL server and is not yet feature--complete. In contrast to our implementation, the JavaScript module does not support the syntax used in the GraphQL specification, i.e. the types have to be specified in a proper format based on JavaScript, while \textit{GraphQL.pl} implements them as a domain--specific language.

Because GraphQL got lots of exposure recently, there are GraphQL implementations in most popular programming languages, including PHP, Java, C/C++, and Haskell. There is a large number of tools to interact with GraphQL or connect it with existing back--end technologies. The \textit{Awesome list of GraphQL and Relay}\footnote{\url{https://github.com/chentsulin/awesome-graphql}} collects a great number of existing GraphQL implementations, tools and related blog posts and conference videos.

\begin{figure}[h]
\begin{center}
  \includegraphics[width=\linewidth]{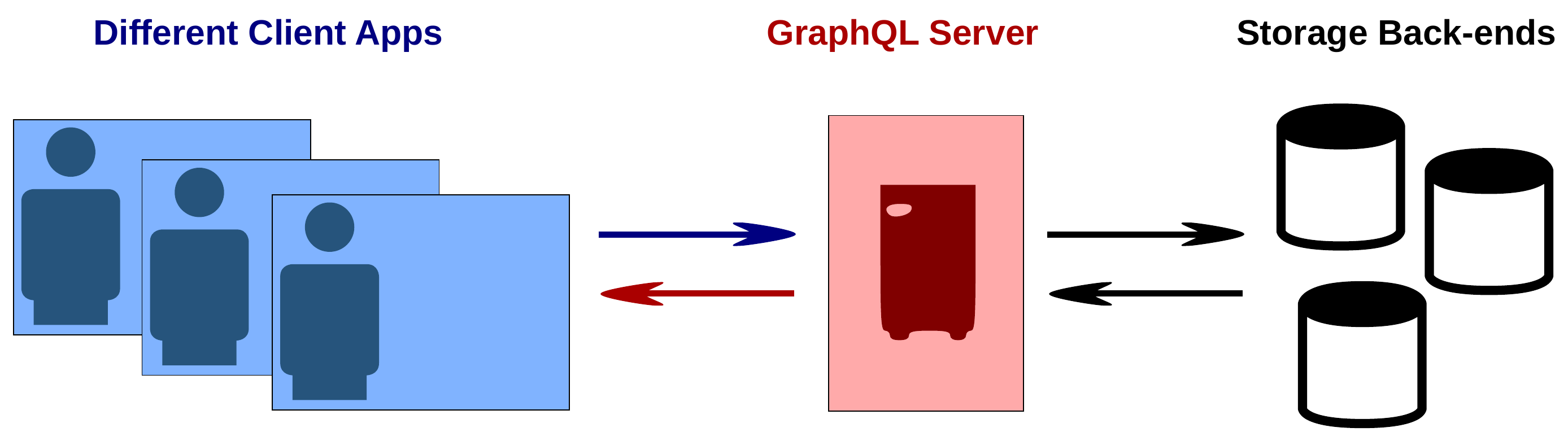}
  \caption{Illustration of the role of the GraphQL Server in a distributed system}
  \label{fig:role}
\end{center}
\end{figure}

The GraphQL server acts as a unified layer for data access and manipulation. In a distributed system, it is located at the same layer like REST, SOAP, and XML--RPC, that means it is used as an abstraction layer to hide the database internals. Similar to these architectural patterns, GraphQL does not support ad--hoc queries. That means that unlike \sql, the general structure of possible queries is defined by the GraphQL server provider in advance. It is not possible to join different entities in a single query, unless their relation has been specified in the GraphQL type definition.

As illustrated in Figure~\ref{fig:role}, client applications written in various programming languages can send requests to the GraphQL server, which validates and executes the query by collecting the information from the various data sources. This protocol is a classical request--response message exchange pattern, with the following two messages:
\begin{itemize}
\item a request, which must be provided in a well--defined query language and send as plain string to the single GraphQL endpoint, and
\item a response of the GraphQL server specified as a \json document.
\end{itemize}

In most applications, complex, structured data is requested from the API. To provide a single entity with all its relations (and possibly their relations, etc.) from a single API endpoint, a GraphQL query is structured hierarchically. The structure of the query represents the data that is expected to be returned. In general, each level of a GraphQL query corresponds to a particular type. They can be nested and even recursive. The result document is a set of entities with their relations specified in the type system. This clarifies the name \texttt{GraphQL}, as these entities and relations can be thought of as a graph.

\begin{figure}[ht]
\begin{minipage}[t]{.5\textwidth}
\begin{lstlisting}
query getAlice {§\mylabel{GraphQL}§
   person(name: "Alice") {
      name
      years: age
      books(favourite: true) {
         # is implicitly a list
         title
         authors {
            name
         }
      }
   }
}
\end{lstlisting}
\end{minipage}
\hfill
\begin{minipage}[t]{.45\textwidth}
\begin{lstlisting}
{ "data": {§\mylabel{Json}§
   "person": {
      "name": "Alice",
      "years": 31,
      "books": [
        {
          "title": "Moby-Dick",
          "authors": [{
            "name": "H. Melville"
          }]
        }
      ]
   } } }
\end{lstlisting}
\end{minipage}
\caption{Example GraphQL query and corresponding result in \json}
\label{code:req_res_example}
\end{figure}

As a motivating example, we will consider a GraphQL server instance to access and manipulate data of persons along with their favourite books. Figure~\ref{code:req_res_example} presents on the left--hand side an example query document to get some basic information about the person named \texttt{Alice}, i.e.~her \texttt{name} and \texttt{age}; the attribute \texttt{age} should be renamed to \texttt{years} in the result set. The query includes her favourite books with their title and authors. The string--based format of the request message presented in the example strictly follows the GraphQL specification. The aim of our work is to implement this format as a domain--specific language in Prolog using DCGs.

On the right--hand side of Figure~\ref{code:req_res_example}, an example result for the query is presented. By definition of the GraphQL specification, the returned data has to be encoded as a \json document. The structure and entities have to follow the request: only keys which were specified in the query are allowed to be part of the result document.

\subsection{Type System}

Being part of the data access and manipulation layer, GraphQL does not support ad--hoc queries like most standard database drivers provide. While with \sql it is possible to dynamically create queries and join multiple tables with almost no restriction, GraphQL queries have to satisfy the given type system defined by the GraphQL server administrator. The entire type system is called the \textit{schema}.

GraphQL's type system is very expressive and supports features like inheritance, interfaces, lists, custom types, enumerated types. By default, every type is nullable, i.e. not every value specified in the type system or query has to be provided.

\begin{figure}[h]
\begin{minipage}[t]{.5\textwidth}
\begin{lstlisting}
type Person {§\mylabel{GraphQL}§
   name:    String!
   age:     Integer
   books(favourite: Boolean): [Book]
   friends: [Person]
}

type Book {
   title:   String!
   authors: [Person]
}
\end{lstlisting}
\end{minipage}
\hfill
\begin{minipage}[t]{.45\textwidth}
\begin{lstlisting}
type Query {§\mylabel{GraphQL}§
   person(name: String!): Person
   book(title: String!):  Book
   books(filter: String): [Book]
}
\end{lstlisting}
\end{minipage}
\caption{Type definitions for the motivating example query of Figure~\ref{code:req_res_example} and GraphQL's \texttt{Query} type}
\label{code:type_definitions}
\end{figure}

In Figure~\ref{code:type_definitions} we present a minimal definition of a type system to satisfy the example query of Figure~\ref{code:req_res_example}. It defines the two types \texttt{Person} and \texttt{Book} as object types, i.e. as a number of field--value--pairs. The fields can have arguments. For example, in order to retrieve the books of a \texttt{Person}, it can be specified to return only favourite books by providing an appropriate \texttt{Boolean} flag.

The two types \texttt{Person} and \texttt{Book} reference each other, i.e.~the underlying model for an author is a person. The type \texttt{Person} is recursive since the \texttt{friends} field returns a list of \texttt{Person}. \texttt{String}, \texttt{Integer}, and \texttt{Boolean} are scalars and among others primitive values defined in the GraphQL specification. The exclamation mark declares values which must not be \texttt{null}, i.e. every \texttt{Person} must specify a \texttt{name} value. Types enclosed by square brackets like \texttt{[Book]} are ordered collections of the given type.

Every GraphQL type system must specify a special root type called \texttt{Query}, which serves as the entry point for the query's validation and execution. As illustrated in Figure~\ref{code:type_definitions}, the fields of this object type can also have parameters, here \texttt{name} and \texttt{title} which are of the type \texttt{String} and must be provided (as denoted by the exclamation mark).

A more detailed introduction to GraphQL's type system, which supports inheritance and composition through fragments, is out of scope of this work. Nevertheless, our presented \textit{GraphQL.pl} system implements the more complex properties of the type system like interfaces, unions, enums, and execution directives.

We want to highlight that the GraphQL specification standardizes only the format of the query document, but not how to denote the type system. The query language is consistent for all GraphQL server implementations and related tools and every GraphQL server has to implement it, independent of the underlying programming language. As opposed to this, the mechanism to specify the type system is currently closely related to the used GraphQL system. For example, in Facebook's reference implementation, the types are defined by calling appropriate JavaScript functions. Instead we propose to use the same domain--specific language as used for the query document. This also eases the usage of the GraphQL specification, because the type examples there are denoted in a format similar to the query document, too.

\subsection{Relation to the Communication Architecture REST}

The type system of GraphQL is similar to the definition of resources in the REST architecture. In both architectural patterns there are no ad--hoc queries, instead the system's types are defined in advance, specifying possible parameters and the output format.

\begin{figure}[h]
\begin{center}
  \includegraphics[width=\linewidth]{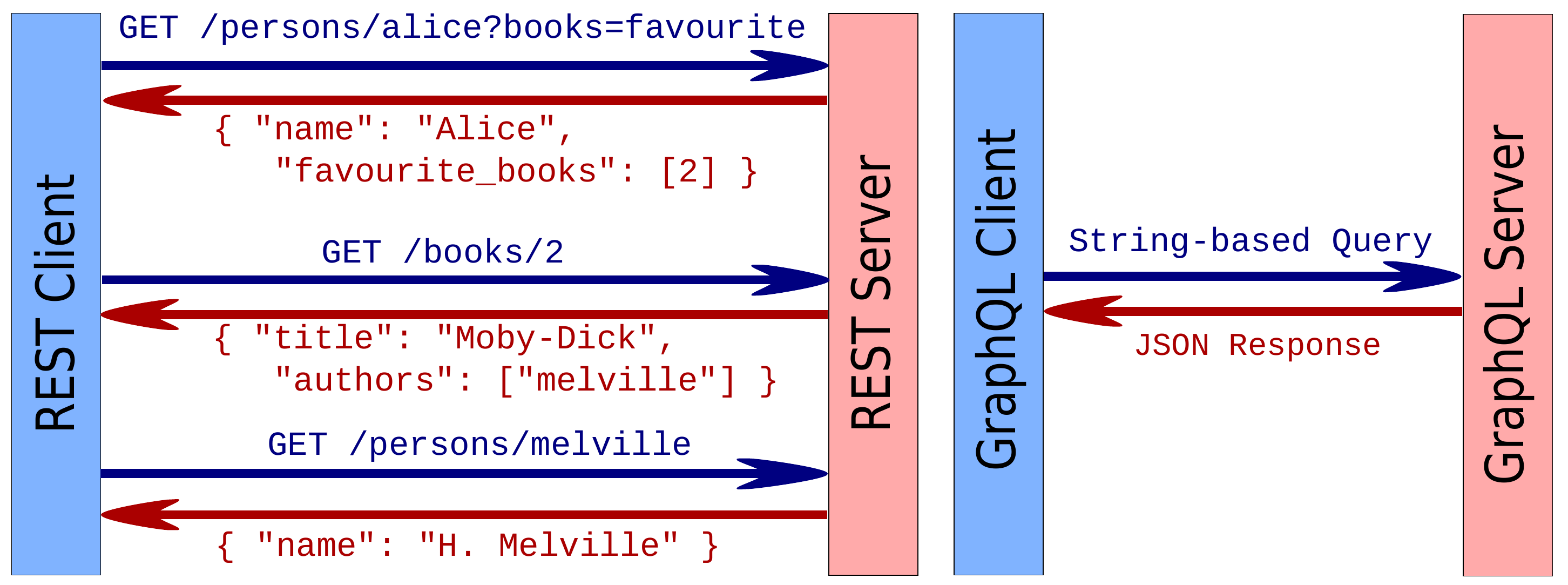}
  \caption{Illustration of the request--response message sequences of a complex query in REST and GraphQL}
  \label{fig:rest-vs-graphql}
\end{center}
\end{figure}

Both GraphQL and REST are independent of the underlying programming language and used da\-ta\-base back--ends. The communication with the client is based on a request--response message exchange pattern. But in contrast to GraphQL, a REST interface in general provides multiple API endpoints, one for each resource. For example, to request complex data consisting of multiple entities and different types, a REST--based protocol has to address multiple resources, resulting in a sequence of round--trip times. The two accessing patterns are illustrated in Figure~\ref{fig:rest-vs-graphql} using the example of requesting a person with its favourite books.

The REST--based data access approach has the need of three round--trip times to collect the required information, which must be transformed into a single object on the client--side. Besides this overhead, REST supports caching and partial documents. The often used transport protocol HTTP adds support for authentication and authorisation, too. These features as well as the freedom of choice with regard to the used data exchange formats are not possible with GraphQL as of yet. In Table \ref{tbl:rest-vs-graphql} we give an overview of the properties of REST and GraphQL.

\begin{table}[h]
\centering
\begin{tabular}{l|c|c}
\textbf{Property}                                                      & \textbf{REST}                                                                        & \textbf{GraphQL}                                                                      \\ \hline\hline
API Endpoints                                                            & multiple                                                                             & single                                                                                \\
\begin{tabular}[c]{@{}l@{}}Message format\\ for queries\end{tabular}   & string--based                                                                         & string                                                                                \\
\begin{tabular}[c]{@{}l@{}}Message format\\ for mutations\end{tabular} & any                                                                                  & string                                                                                \\
\begin{tabular}[c]{@{}l@{}}Message format\\ for response\end{tabular}  & \begin{tabular}[c]{@{}c@{}}any,\\ especially hypermedia\end{tabular}                 & \json                                                                                  \\ \hline
Type System                                                            & \begin{tabular}[c]{@{}c@{}}weakly typed,\\ no machine--readable metadata\end{tabular} & \begin{tabular}[c]{@{}c@{}}strongly typed,\\ meta data for introspection\end{tabular} \\
Built--in features                                                      & cacheable                                                                            & query validation, introspection                                                      
\end{tabular}
\vspace*{0.2cm}
\caption{Comparison of REST and GraphQL}
\label{tbl:rest-vs-graphql}
\end{table}

\section{Definition of GraphQL as a DSL}
\label{sec:dsl}

The aim of our implementation is to support the GraphQL syntax presented in its specification to ease the use of the given examples with our Prolog implementation. The syntax for query documents is well--defined in the specification. However, the format to declare types\footnote{The GraphQL types are declared when setting up a GraphQL server, generally by the database administrator.} is not standardized at all and might therefore be different in GraphQL implementations of other programming languages. We choose to follow the format given for the code examples in the GraphQL specification. In this way, the code snippets mentioned in the documentation can easily be used to create a standalone GraphQL server.

To support the GraphQL syntax in \textit{GraphQL.pl}, we define an appropriate domain--specific language~(DSL), for what Prolog is very suitable. Simple languages can be modelled directly using Prolog syntax only, i.e. by defining terms and declaring appropriate operators. For instance, the \json data format as presented on the right--hand side of Figure~\ref{code:req_res_example} is valid Prolog syntax, although a shorter notation is feasible using Prolog atoms.

In a similar way, a subset of the GraphQL language could be represented by Prolog terms with only minimal modifications, i.e. the language could be implemented as an internal DSL. As a result, it would be possible to use the GraphQL examples directly within Prolog, without having the need to write a parser first. The definition as an internal DSL would also profit from tools actually written for Prolog.

We want to discuss the definition of GraphQL as an internal DSL in Prolog based on the declaration of the type \texttt{Person} as presented in Figure \ref{code:type_definitions}. By defining the \texttt{(!)/1} suffix operator, the declaration of non--nullable types is syntactically valid. Similarly, we define \texttt{type/1} as a prefix operator. Since whitespaces and commas are insignificant for GraphQL and type names can start with a lowercase letter, the \texttt{Person} type definition of Figure~\ref{code:type_definitions} could be expressed by the term presented in Figure~\ref{code:internal_dsl}, which is valid syntax of standard Prolog.

\begin{figure}[h]
\begin{lstlisting}
:- op(800, xf, !).§\mylabel{Prolog}§
:- op(650, fx, type).
?- Person = type person(
               name: string!,
               age: integer,
               friends: [person],
               books: [book] ).
\end{lstlisting}
\caption{Operator definitions to define GraphQL as an internal DSL}
\label{code:internal_dsl}
\end{figure}

The notation of Figure~\ref{code:internal_dsl} almost follows the GraphQL specification, only the curly braces had to be replaced to achieve a valid Prolog term\footnote{Since the introduction of dicts in SWI--Prolog 7, \texttt{person\{\}} would be syntactically correct, i.e. it is not necessary to replace the curly braces. We present dicts in Section \ref{sec:implementation:dicts}.}. Nevertheless, this approach of defining the GraphQL type specification as an internal DSL in Prolog is very restrictive:
\begin{itemize}
\item Types have to start with a lowercase letter to not create a variable. Although this is allowed according to the GraphQL specification, it is convention to start type names with an uppercase letter. Especially it is not possible to reference built--in types, since \texttt{String}, \texttt{Integer} etc. all start with an uppercase letter.
\item Because we replaced curly with normal braces, query documents could contain compounds with two pairs of brackets: one for the arguments, another for nested items. In the query document of Figure~\ref{code:req_res_example}, this results in \texttt{person(name:\,'Alice')(name,\,years:\,age,\,...)}, which is not valid in Prolog\footnote{Even using dicts in SWI--Prolog 7, i.e. to use the term \texttt{person(name:\,'Alice')\{name:\,\_,\,years:\,age,\,...\}}, this would not be valid, since a dict's tag must be either a variable or atom.}.
\end{itemize}

Because of these restrictions and since \textit{GraphQL.pl} shall completely support the syntax given in the GraphQL specification, the definition of the language as an external DSL might be more reasonable. In the following we will develop a way of embedding GraphQL in SWI--Prolog 7 using quasi quotations.

\section{Implementation Using Features of SWI--Prolog 7}
\label{sec:implementation}

We have implemented the \textit{GraphQL.pl} system in SWI--Prolog~\cite{wielemaker2003overview}, using features recently added to SWI--Prolog of version 7~\cite{wielemaker2014swi}. Our implementation makes great use of quasi quotations, definite clause grammars, and dicts.

\subsection{Quasi Quotations}

For more complex languages or if Prolog's syntax is too restrictive, other mechanisms for the definition of external DSLs have been discussed~\cite{wielemaker2012syntactic}. Since Prolog has poor support for quoting long text fragments, quasi quotations have been introduced to SWI--Prolog in 2013~\cite{wielemaker2013s}. They are inspired by Haskell and in similar form also part of other programming languages like JavaScript (called \textit{tagged template strings}). Quasi quotations in SWI--Prolog have been used to implement external languages like \html, \sql, and \sparql.

The basic form of a quasi quotation is of the following form, where \texttt{Tag} is a callable Prolog predicate with an arity of 4 and \texttt{Content} is a string:

\begin{quote}
\verb@{|Tag||Content|}@\hspace*{1cm} e.g.\hspace*{1cm}\verb@{|html||<p>Hello World!</p>|}@
\end{quote}

\noindent For better readability in this paper, we will enclose the \texttt{Tag} by whitespaces, which is still valid SWI--Prolog 7 syntax. The quasi quotation presented before invokes the following call by the Prolog parser. Because quasi quotations are resolved using Prolog's term expansion mechanism, the following call is invoked at compile time:

\begin{quote}
\begin{BVerbatim}
call(+TagName, +ContentHandle, +SyntaxArgs, +Vars, -Result)
\end{BVerbatim}
\end{quote}

The quasi quotation's \texttt{Tag} can be any term with the functor \texttt{TagName/TagArity}. The \texttt{Tag}'s arguments are passed as a list \texttt{SyntaxArgs}\footnote{This can also be expressed as: \texttt{Tag\,=..\,[TagName|SyntaxArgs],\,length(SyntaxArgs, TagArity)}.}, which guarantees that the \texttt{TagName} predicate is always of arity 4. The \texttt{Vars} argument provides access to variables of the quasi quotation's context. \texttt{ContentHandle} is an opaque term that carries the content of the quasi quoted text and position information about the source code. It is usually passed to \texttt{with\_quasi\_quotation\_input/3} or \texttt{phrase\_from\_quasi\_quotation/2}. While the first predicate creates a stream, we are using \texttt{phrase\_from\_quasi\_quotation/2}, which parses the enclosed string according to a given DCG. For instance, Figure~\ref{code:quasi_quotation_type} illustrates the definition of the \texttt{\{| type ||~...~|\}} quasi quotation, which processes the given \texttt{Content} by using the non--terminal \texttt{dcg\_type} of the DCG.

\begin{figure}[h]
\begin{lstlisting}
:- use_module(library(quasi_quotations)).§\mylabel{Prolog}§
:- quasi_quotation_syntax(type).

type(ContentHandle, SyntaxArgs, Vars, Result) :-
   phrase_from_quasi_quotation(dcg_type(Vars, Result), ContentHandle).
  
dcg_type(Vars, Result) --> ...  % DCG definitions
\end{lstlisting}
\caption{Definition of the \texttt{type} quasi quotation}
\label{code:quasi_quotation_type}
\end{figure}

In our implementation, we use quasi quotations for being able to embed the query document as well as the type definitions in the Prolog source code. In order to define types and the overall schema, \textit{GraphQL.pl} provides several quasi quotation tags, for example \texttt{type/4} and \texttt{schema/4}. Figure~\ref{code:quasi_quotation_syntax} gives an example on how to use the quasi quotations to define types and a schema. All user--defined type names starting with an uppercase letter create corresponding Prolog variables, i.e. the quasi quotations in Figure~\ref{code:quasi_quotation_syntax} make the variables \texttt{Person}, \texttt{Book} and \texttt{Query} available. In this way, it is possible to reference the \texttt{Book} type in the type definition of \texttt{Person} and vice--versa. In contrast, the built-in types like \texttt{Integer} create atoms beginning with a lowercase letter, i.e. \texttt{integer} etc.

Besides these tags, we define quasi quotations for simple, text--based specifications of \texttt{enum} types and \texttt{interface} declarations. The query and mutation documents can be specified with a quasi quotation as well, although in general they are given as strings as part of the GraphQL request sent to the server.

\begin{figure}[h]
\begin{lstlisting}
:- use_module(library(graphql), [type/4, schema/4]).§\mylabel{Prolog}§
example_schema(S) :-
   Person = {| type ||
      name:                      String!
      age:                       Integer
      books(favourite: Boolean): [Book]
      friends:                   [Person]
   |},
   S = {| schema ||
      type Query {
         person(name: String!): Person
         book(title: String!):  Book
         books(filter: String): [Book]
      }
      type Book { ... }
   |}.
\end{lstlisting}
\caption{Definition of the type \texttt{Person} and schema \texttt{S} using quasi quotations}
\label{code:quasi_quotation_syntax}
\end{figure}

\subsection{Definite Clause Grammars}

The GraphQL specification provides a grammar to define the format of a query document. It is the basis for the DCGs~\cite{pereira1980definite} used in our implementation to parse the query document. For parsing the type definitions, the same mechanism is used.

In Figure~\ref{query_dcg} we present the simplified grammar to parse the \texttt{type} quasi quotation. It is used by \linebreak\texttt{phrase\_from\_quasi\_quotation/2} as shown in Figure~\ref{code:quasi_quotation_type}. For the sake of simplicity, we omit the generation of the internal representation \texttt{Result}, and also the handling of whitespaces and the specification of minor DCG rules. Using this grammar, the type definition specified in Figure~\ref{code:quasi_quotation_syntax} can be processed.

\begin{figure}[h]
\begin{lstlisting}
dcg_type(_Vars, _Result) -->       % ignore for simplicity§\mylabel{Prolog}§
   list_of_type_definition.        % resolves to 'type_definition'
type_definition -->
   name,                           % the type's name in the object
   ( arguments | "" ),             % optional, e.g. favourite: Boolean
   ":",
   type.
type -->
   ( named_type                    % e.g. Integer
   | list_type                     % e.g. [Book]
   | non_null_type ).              % e.g. String!
named_type -->
   ( primitive_type                % built-in type of GraphQL
   | prolog_var_name ).            % part of library(dcg/basics)
list_type --> "[", type, "]".
non_null_type -->
   ( named_type | list_type ),
   "!".
\end{lstlisting}
\caption{Extract of the DCG to parse the \texttt{type} quasi quotation}
\label{query_dcg}
\end{figure}

Most of the grammar specified in the GraphQL specification for query documents has been translated into a DCG, resulting in more than 50 non--terminals just to parse queries. The entire DCG rule base to parse the type system as well is implemented in more than 600 lines of code with more than\linebreak 80 non--terminals. Their names are based on the grammars presented in the GraphQL specification. However, to define sequences of elements, the DCG easily gets bloated by helping rules. Because the Extended Definite Clause Grammars~\cite{schneiker2009declarative} are also part of the \ddk, we intend to adopt \edcg rules for writing more powerful grammars in a simpler way in the future.

\subsection{Internal Representation Using Dicts}
\label{sec:implementation:dicts}

The DCG is used to parse the given quasi quotations of type definitions and query documents. Along with this, we create an abstract syntax tree (AST) as an internal representation. We use SWI--Prolog's dicts to create nested terms holding the given information.

With SWI--Prolog 7, dicts were introduced as a new data type for named key--value associations~\cite{wielemaker2014swi}. In previous versions this was often realised using a list of \texttt{Key=Value} pairs. For example \json documents were encoded as what follows:

\begin{quote}
\begin{BVerbatim}
json([ name="Alice", years=31, ... ])
\end{BVerbatim}
\end{quote}

This notation does not ensure the uniqueness of the keys and does not provide a short notation to directly access a child node. The new syntax of dicts resembles the one of \json and should be more familiar for developers of other programming languages. It is of the following form\footnote{In the following, our code examples will include both quasi quotations and dicts, which might confuse the reader. Although their visual appearance is similar, they can easily be distinguished by their start token:
\begin{itemize}
\item quasi quotations are opened by \texttt{\{|},
\item dicts with \texttt{Tag\{}.
\end{itemize}}:

\begin{quote}
\begin{BVerbatim}
Tag{ Key1: Value1, Key2: Value2, ... }
\end{BVerbatim}
\end{quote}

The \texttt{Tag} is either an atom or a variable. We also make use of the anonymous dict \texttt{\_\{...\}}, where \texttt{Tag} is the anonymous variable. For example, the type information created by the \texttt{type} quasi quotation are encoded by a dict of the following form:

\begin{quote}
\begin{BVerbatim}
object{
   fields: _{
      FieldName1: field{
         type: Type1,
         resolver: Resolver1      
      }, ... },
   resolve: Resolver }
\end{BVerbatim}
\end{quote}

This nested \texttt{object\{\}} dict contains a collection of type definitions as the \texttt{fields} entry. It corresponds to the \texttt{list\_of\_type\_definitions} DCG rule of Figure~\ref{query_dcg}, i.e.~it has a \texttt{FieldName} for every \texttt{type\_definition}. The \texttt{resolve} entries are automatically created for every type and field. They are used to specify how to retrieve the data for this component and are unbound at first. In Section~\ref{sec:example} we introduce their usage in detail.

Dicts can be unified following the standard symmetric Prolog unification rules, although the unification will fail if both dicts do not contain the same set of keys:

\begin{quote}
\begin{BVerbatim}
?- p{ a: 1, b: 2 } = P{ a: A, b: B }.
P = p, A = 1, B = 2.

?- p{ a: 1, b: 2 } = P{ a: A }.
false.
\end{BVerbatim}
\end{quote}

Besides this, there are two methods to access a single value of the \texttt{Key} in the dict. With \texttt{Dict.Key} the related value is retrieved. Since this \textit{dot notation} will throw an error if the given \texttt{Key} does not appear in the \texttt{Dict}, we prefer using \texttt{Dict.get(Key)}, which instead silently fails for missing keys and can therefore be used as a pre--condition in a Prolog rule.

After parsing the quasi quotations using DCGs, the GraphQL schema and types are represented by dicts. The AST generation was omitted in the code example of Figure~\ref{query_dcg} for the sake of simplicity. In Figure~\ref{type_ast} we present an extract of the generated dicts.

\begin{figure}[h]
\begin{lstlisting}
example_schema(S) :-§\mylabel{Prolog}§
   Person = object{
      fields: _{
         name: field{ type: string, nonNull: true, resolve: _ },
         age: field{ type: integer, resolve: _ },
         books: list{
            arguments: _{ favourite: field{ type: boolean } },
            kind: field{ type: Book }, resolve: _ },
         friends: list{ kind: field{ type: Person }, resolve: _ } },
      resolve: _ },
   S = schema{
      query: object{
         fields: _{
            person: field{ type: Person, ... },
            book: field{ ... }, books: field{ ... } }, ... } }.
\end{lstlisting}
\caption{Extracts of the generated dicts for the type and schema definitions of Figure~\ref{code:quasi_quotation_syntax}}
\label{type_ast}
\end{figure}

\noindent
In SWI--Prolog without occurs check, a call \texttt{Person\,=\,object\{\,...\,Person\,...\,\}} does not raise an error. Instead, a corresponding cyclic substitution is created. Later, we will select the types of sub fields of \texttt{Person} using the dot notation, e.g., \texttt{T = Person.fields.friends.kind.type}. Since only GraphQL's types can be cyclic but not the queries, this will not lead to non-termination.

The AST for the query document is of an equivalent form. In Figure~\ref{type_ast} we omitted additional internal properties used for introspection. For example it is possible to give each type and field a textual description which can be accessed with the help of the built--in \texttt{\_\_type} query.

\section{Query Execution Using Type and Field Resolvers}
\label{sec:example}

In this section, we present a working example for \textit{GraphQL.pl}, which illustrates the query execution and validation. For data storage, we use Prolog facts \texttt{person/6} and \texttt{book/3} as presented in Figure~\ref{code:graphql_example_facts}. Their data correspond to the type definitions of Figure~\ref{code:type_definitions}. To return either all or only the person's favourite books based on the \texttt{favourite} argument we use an additional \texttt{Favs} data entry. The example can easily be adapted to request the data from external databases with the help of \sql or from text--based formats like \xml and \csv.

\begin{figure}[h]
\begin{lstlisting}
% person(Id, Name, Age, Books, Favs, Friends)§\mylabel{Prolog}§
person(1, 'Alice', 31, [1, 2], [2], [2]).
person(2, 'Bob', 42, [2], [2], [1, 3]).
person(3, 'H. Melville', 72, [1], [], []).
person(4, 'D. Defoe', 71, [], [], []).

% book(Id, Title, Authors)
book(1, 'Robinson Crusoe', [4]).
book(2, 'Moby-Dick', [3]).
\end{lstlisting}
\caption{Example fact base}
\label{code:graphql_example_facts}
\end{figure}

In order to execute a query document, both the AST of the query and of the type system are traversed simultaneously in a top--down approach. Beginning with the root type \texttt{Query} specified in the GraphQL schema, the \textit{GraphQL.pl} system searches for the requested fields in the type definitions. To get the appropriate value, for every type of the schema a \texttt{resolve/5} predicate has to be defined, which we call the \textit{resolver}. It is used to generate the resulting dict for a specific type, which is the basis for the \json document returned to the client. A resolver is of the following form:

\begin{quote}
\begin{BVerbatim}
resolve(+Id, +Parent, +Args, +AST, -Result)
\end{BVerbatim}
\end{quote}

It is called every time the \textit{GraphQL.pl} system has to retrieve the value of a field in the query document. This \texttt{Result}, which must be returned as a dict, is made up based on the given context information: \texttt{Id} is a unique identifier which might be used to retrieve a single data set; \texttt{Args} is a dict with all the arguments specified for the current field and might be empty, i.e. \texttt{\_\{\}}; \texttt{AST} provides the static type information of the currently examined GraphQL type. The \texttt{Parent} argument passes the calculated result of the superior document level. In this way it is possible to, e.g., create a particular resolver for the \texttt{books} field in the \texttt{Person} type of Figure~\ref{code:type_definitions}, which has access to the calculated answer dict of the person. This technique is used to refine or complete the answer of the parent's level.

The resolver's arguments might not be bound on every predicate call, for example, there is no \texttt{Parent} information for the query's root type. In most cases no unique identifier \texttt{Id} is specified, instead the \texttt{Result} has to be shaped based on the \texttt{Parent} and \texttt{Args} context information.

\begin{figure}[h]
\begin{lstlisting}
resolve_p(Id, _Parent, Args, _AST, Result) :-§\mylabel{Prolog}§
    ( Name = Args.get(name)
    ; nonvar(Id) ),
    person(Id, Name, Age, Fri, All, Favs),
    Result = _{ name:Name, age:Age, friends:Fri, books:(All, Favs) }.

resolve_b(Id, _Parent, Args, _AST, Result) :-
    ( book(Id, Authors, Args.get(title))
    ; nonvar(Id), book(Id, Authors, Title) ),
    Result = _{ title:Title, authors:Authors }.
    
resolve_p_b(_Id, Parent, Args, _AST, Result) :-
    (All, Favs) = Parent.books,
    ( Args.get(favourite) = true ->
      Result = Favs
    ; Result = All ).
\end{lstlisting}
\caption{Definition of the resolver predicates}
\label{code:graphql_example_resolvers}
\end{figure}

In Figure~\ref{code:graphql_example_resolvers} we define the resolvers for our usage example\footnote{To use the \textit{GraphQL.pl} system, the user has to implement a resolver for every GraphQL type, so the system is able to retrieve the requested data according to these resolver predicates. Since the resolvers are explicitly assigned to the appropriate types later, the predicate's name is not significant. We choose to call them using the type's first letter, i.e.~\texttt{resolve\_p} is the resolver for the type \texttt{Person}.}, \texttt{resolve\_p/5}, \texttt{resolve\_b/5}, and \texttt{resolve\_p\_b/5}. They are required to calculate the answer dicts based on the given fact base:

\begin{itemize}
\item \texttt{resolve\_p/5} is the used resolver for the type \texttt{Person}. It collects the person's information by its name, provided in the \texttt{Args} dict. If otherwise no \texttt{name} argument is specified, i.e. \texttt{Args.get(name)} fails, the appropriate data entry is retrieved by the given identifier \texttt{Id}.
\item \texttt{resolve\_b/5} retrieves the data for the type \texttt{Book} and similarly depends either on the book's title or its identifier.
\item In addition to \texttt{resolve\_p/5} we define the predicate \texttt{resolve\_p\_b/5}, which is used to only determine the returned value for the \texttt{books} field of a person.
\end{itemize}

The result of a requested \texttt{Person} type depends on the value of the boolean flag \texttt{favourite} of the \texttt{books(favourite:\,Boolean)} selector in the query: if it is \texttt{true}, only the person's favourite books should be listed; otherwise all. Therefore we determine the value of the \texttt{books} field in two steps, using the predicates mentioned before:

\begin{enumerate}
\item When determining the field--value associations for the \texttt{Person} object type by calling the \linebreak\texttt{resolve\_p/5} predicate, the value for the \texttt{books} field is set to the tuple \texttt{(All,Favs)}.
\item Because the fields are resolved in a top--down approach, \textit{GraphQL.pl} now takes into account possibly defined resolvers of the next level's fields. We use the \texttt{resolve\_p\_b/5} resolver as a refinement of the \texttt{books} field. It returns either the \texttt{All} or \texttt{Favs} tuple element provided in \texttt{Parent.books}, depending on the \texttt{favourite:\,Boolean} flag given as \texttt{Args.favourite} property.
\end{enumerate}

As a last step for the \textit{GraphQL.pl} server declaration, the resolvers have to be connected to the appropriate types and fields. Every type and field generated using \textit{GraphQL.pl}'s quasi quotations has a special \texttt{resolve} key which is unbound at first. It can be instantiated using the dot notation introduced in SWI--Prolog 7 for dicts. In our example, the type resolver for \texttt{Person} is declared by \linebreak\texttt{Person.resolve = resolve\_p}, as presented in Figure~\ref{code:graphql_example_query}. The resolver for the person's \texttt{books} field is declared by \texttt{Person.fields.books.resolve\,=\,resolve\_p\_b}.

\begin{figure}[h]
\begin{lstlisting}
:- use_module(library(graphql), [query/3]).§\mylabel{Prolog}§
?- Schema = {| schema ||
      type Query { ... }
      type Book { ... } |},
   Person = {| type || ... |},
   Person.resolve = resolve_p,
   Person.fields.books.resolve = resolve_p_b,
   Book.resolve = resolve_b,
   Query = {| query ||
      person(name: "Alice") {
         name, years: age
         books(favourite: true) { title, authors { name } } } |},
   graphql:query(Schema, Query, Result).
\end{lstlisting}
\caption{Declaration of the schema and the resolvers with following query execution}
\label{code:graphql_example_query}
\end{figure}

In a real--world application, the queries are raised by a remote client. The GraphQL server gets the query as string, runs the execution and sends back the response as a \json document. To complete our given example, we directly call the query using the predicate \texttt{query/3} of \textit{GraphQL.pl}, where \texttt{Schema} and \texttt{Query} are dicts generated from the corresponding quasi quotations:

\begin{quote}
\begin{BVerbatim}
graphql:query(+Schema, +Query, -Result)
\end{BVerbatim}
\end{quote}

\noindent
\texttt{query/3} creates a result according to the type system and the resolver predicates
and returns it as a dict, which could be directly sent to the client as a \json document. In Figure~\ref{code:graphql_example_query} we give an example on specifying and executing a query given as a quasi quotation. \texttt{Result} is a dict representing the \json output already stated in Figure~\ref{code:req_res_example}. Because dicts are the preferred format to specify \json since SWI--Prolog~7, they can be, e.g., printed by using the built--in \texttt{json\_write\_dict/3}. An example result is\footnote{It is worth noting that although the type system given in Figure~\ref{code:type_definitions} defined \texttt{books} as a list of \texttt{Book}, i.e. \texttt{book:\,[Book]}, this is neither expressed in the string representation of the query nor its AST. But because this information can be deducted from the type system, the correct result document is calculated and the query is valid and unambiguous.}:

\begin{quote}
\begin{BVerbatim}
Result = _{ data: _{
   person: _{ name: 'Alice', years: 31,
      books: [ _{ title: 'Moby-Dick',
         authors: [ _{ name: 'H. Melville' } ] } ] } } }
\end{BVerbatim}
\end{quote}

\section{Conclusions and Future Work}
\label{sec:conclusion}

In this paper, we have introduced the language GraphQL, an application layer used for data queries and manipulations developed by Facebook, and presented a comparison to REST, a well--established architectural pattern for distributed systems. We have implemented a GraphQL server in SWI--Prolog 7, called \textit{GraphQL.pl}. We used features added to SWI--Prolog very recently in version 7, resulting in short and very readable source code. The \textit{GraphQL.pl} system is an example for these features to underline their need and production--ready state in a real application.

For being able to specify the query and type system in the same way it is used in the GraphQL specification, the text--based format has been implemented as an external domain--specific language in Prolog. Along with the used syntactic elements -- quasi quotations, definite clause grammars and dicts~--, this allows for the specification of a GraphQL server even for developers not yet familiar with Prolog. In the GraphQL reference implementation, similar syntactic elements have been used in JavaScript: tagged template strings to embed the type system DSL, and \json for the returned objects.

As the \textit{GraphQL.pl} system has been developed in a test--driven approach, it also provides a test framework with a large number of use--cases of the implemented domain--specific language. The entire implementation is available online at \url{https://github.com/fnogatz/GraphQL.pl}.

In the future, we intend to connect \textit{GraphQL.pl} with more back--end technologies, e.g. \sql, \xml and REST APIs. Because our deductive database system \ddbase~\cite{seipel2015knowledge} already supports various, heterogeneous data sources, we intend to use GraphQL as an additional unified query language. Using the data's meta information collected by \ddbase, one could target reasoning about the data's format in order to improve GraphQL's introspection system or automatically generate the type system.

Because -- in contrast to a REST approach -- GraphQL is based only on a single request--response cycle, all the fields requested by the client are known at query execution, we intend to incorporate optimising strategies in the resolvers. In this way, it would be possible to use the abstraction layer introduced by GraphQL for query optimisations for existing query languages like \sql, too.

Besides for data access, GraphQL specifies also the syntax for mutating data sets. The \texttt{mutation} operation is not implemented in \textit{GraphQL.pl} so far. As a proof--of--concept, we intend to connect GraphQL mutations with FN--Query~\cite{seipel2002processing}, which allows the data access and manipulation of semi--structured data in XML documents.

\nocite{*}
\bibliographystyle{eptcs}
\bibliography{Literature}
\end{document}

%% file: SeipelStyles.tex
\newcommand{\xml}{\textsc{Xml}\xspace}

\newcommand{\ddbase}{\textsc{DDbase}\xspace}
\newcommand{\ddk}{\textsc{Ddk}\xspace}

\newcommand{\eat}[1]{}

\newenvironment{dqnarray*}{%
   \begin{quote}
       \( \begin{array}{lcl}}{\end{array} \)
   \end{quote} }
\newenvironment{mqnarray*}{%
   \vspace*{2mm}
   \begin{quote}
      \( \begin{array}{l}}{\end{array} \)
   \end{quote} }

  \newcommand{\bi}{\begin{itemize}}
  \newcommand{\ei}{\end{itemize}}

  \newcommand{\sql}{\textsc{Sql}\xspace}
  \newcommand{\html}{\textsc{Html}\xspace}
  \newcommand{\sparql}{\textsc{Sparql}\xspace}
  \newcommand{\json}{\textsc{Json}\xspace}
  \newcommand{\csv}{\textsc{Csv}\xspace}
  
  \newcommand{\edcg}{\textsc{EDcg}\xspace}

